\documentclass[a4paper]{spie}  

\usepackage{amsmath,amsfonts,amssymb}
\usepackage{graphicx}
\usepackage[colorlinks=true, allcolors=blue]{hyperref}
\usepackage{amsmath,amsfonts,amssymb,mathtools,float,textcomp}
\usepackage{verbatim}
\usepackage[table,xcdraw]{xcolor}
\usepackage{amsmath}
\usepackage{array}
\usepackage{geometry}

\title{The multi$-$physics analysis and design of CUSP, a two CubeSat constellation for Space Weather and Solar flares X-ray polarimetry}

\author[a,g]{Giovanni Lombardi}
\author[a]{Sergio Fabiani}
\author[a]{Ettore Del Monte}
\author[a]{Enrico Costa}
\author[a]{Paolo Soffitta}
\author[a]{Fabio Muleri}
\author[d]{Ilaria Baffo}
\author[g]{Marco E. Biancolini}
\author[b]{Sergio Bonomo}
\author[j]{Daniele Brienza}
\author[f]{Riccardo Campana}
\author[h]{Mauro Centrone}
\author[e]{Gessica Contini}
\author[b]{Giovanni Cucinella}
\author[c]{Andrea Curatolo}
\author[a]{Nicolas De Angelis}
\author[f]{Giovanni De Cesare}
\author[e]{Andrea Del Re}
\author[a]{Sergio Di Cosimo}
\author[b]{Simone Di Filippo}
\author[a]{Alessandro Di Marco}
\author[g]{Emanuele Di Meo}
\author[a]{Giuseppe Di Persio}
\author[j]{Immacolata Donnarumma }
\author[d]{Pierluigi Fanelli}
\author[e]{Paolo Leonetti}
\author[c]{Alfredo Locarini}
\author[a]{Pasqualino Loffredo}
\author[g]{Andrea Lopez}
\author[i]{Gabriele Minervini}
\author[c]{Dario Modenini}
\author[j]{Silvia Natalucci}
\author[b]{Andrea Negri}
\author[b]{Massimo Perelli}
\author[a]{Monia Rossi}
\author[a]{Alda Rubini}
\author[a]{Emanuele Scalise}
\author[j]{Andrea Terracciano}
\author[c]{Paolo Tortora}
\author[j]{Emanuele Zaccagnino}
\author[e]{Alessandro Zambardi}

\affil[a]{INAF-IAPS, via del Fosso del Cavaliere 100, 00133, Roma, Italy}
\affil[b]{IMT s.r.l., via Carlo Bartolomeo Piazza 30, 00161 Rome, Italy}
\affil[c]{Alma Mater Studiorum Universit\'a di Bologna - Department of Industrial Engineering and Interdepartmental Center for Industrial Aerospace Research, Via Fontanelle 40, 47121 Forl\'i, Italy}
\affil[d]{DEIM, Universita\'a degli studi della Tuscia, Largo dell’Universit\'a, 01100 Viterbo, Italy}
\affil[e]{SCAI Connect s.r.l., Via Vincenzo Lamaro 51, 00173 Roma, Italy}
\affil[f]{INAF-OAS Bologna, via Pierobetti 93/3, 40129 Bologna, Italy}
\affil[g]{Dipartimento di Ingegneria dell'Impresa "Mario Lucenti", Università degli Studi di Roma "Tor Vergata", Via Cracovia 50, 00133 Roma, Italy}
\affil[h]{INAF-OAR, Via Frascati 33, 00040, Monte Porzio Catone, Italy}
\affil[i]{INAF Headquarters, Viale del Parco Mellini 84, 00136,  Roma, Italy}
\affil[j]{ASI, Via del Politecnico snc 00133 – Roma, Italy}

\authorinfo{Further author information: (Send correspondence to G. Lombardi)\\G. Lombardi: E-mail: giovanni.lombardi@inaf.it}

\begin{document} 
\maketitle
\begin{abstract}
The CUbesat Solar Polarimeter (CUSP) project aims to develop a constellation of two CubeSats orbiting the Earth to measure the linear polarization of solar flares in the hard X-ray band by means of a Compton scattering polarimeter on board of each satellite. CUSP will allow to study the magnetic reconnection and particle acceleration in the flaring magnetic structures. CUSP is a project approved for a Phase B study by the Italian Space Agency in the framework of the Alcor program aimed to develop CubeSat technologies and missions.
In this paper we describe the a method for a multi-physical simulation analysis while analyzing some possible design optimization of the payload design solutions adopted. In particular, we report the mechanical design for each structural component, the results of static and dynamic finite element analysis, the preliminary thermo-mechanical analysis for two specific thermal cases (hot and cold orbit) and a topological optimization of the interface between the platform and the payload. 
\end{abstract}

\keywords{CUSP, space weather, solar flare, mechanical design, multi$-$physical analysis, payload}

\section{INTRODUCTION}
\label{sec:intro}

The CUbesat Solar Polarimeter (CUSP)\cite{fabiani1} project is a CubeSat mission orbiting the Earth aimed to measure the linear polarization of solar flares in the hard X$-$ray band (25$-$100 keV energy band) by means of a Compton scattering polarimeter. CUSP will allow to study the magnetic reconnection and particle acceleration in the flaring magnetic structures of our star. CUSP is a project in the framework of the Alcor Program of the Italian Space Agency aimed to develop innovative CubeSat missions. It is approved for a Phase B to start in 2024\cite{fabiani2024cusp}.
The activity presented is aimed to develop a method for a multi$-$physical simulation analysis while analyzing some possible design optimization of the payload.

\section{WORKFLOW AND SOFTWARE}
The multifaceted design of a Cubesat mission presents various engineering challenges.\\ Specifically, one must contend with minimal geometric dimensions, reduced weight for each unit of the platform, and stringent scientific requirements. It is essential to generate and refine a mechanical baseline for the core of the instrument, the sensitive part of CUSP, while harmonizing the different mission requirements. Starting from the phase A design we implemented a method to assess each mechanical part and the internal and external interfaces of the payload.\\
Finally, an iterative optimization preliminary method was developed and is reported here, encompassing various physical aspects (mainly mechanical and thermal) and potentially leading to the refinement of the mission design.\\
The software used for this project are the following:
\begin{itemize}
   \item \textbf{SolidWorks}\\
   SolidWorks is a three$-$dimensional parametric design software, produced and marketed by Dassault Systèmes. Useful for the design of mechanical devices, even complex ones, it involves the creation of 2D and 3D drawings of solids and surfaces, through a parametric geometric system \cite{dass}.
   \item \textbf{ANSYS SpaceClaim}\\
   ANSYS Space Claim is a parametric solid modeling software available in the ANSYS package that operates mainly according to the direct modeling logic, thanks to which it is possible to define the construction and modification operations of the desired geometry, without the need to define interdependent features.
   \item \textbf{ANSYS Meshing}\\
   ANSYS Meshing is the software responsible for generating the calculation grid. This process is automated in its simplest application although it is possible to add control settings to act directly on those parameters that define the mesh in terms of morphology, topology and dimensions of the constituent cells.
\item \textbf{ANSYS Workbench}\\
    The Workbench platform is the tool designed to put in communication the various modeling and numerical analysis software that fall within the ANSYS landscape, facilitating the management of the project that is articulated there through various work steps. Its key feature is to allow the transfer and sharing of input and output data not only between the different tools used in pre and post processing, but also between multiple simulation models (which allows for the easy implementation of sophisticated multi$-$physical simulations) \cite{ansys}.
\item \textbf{Systema Thermica}\\
    Thermica is a sophisticated thermal analysis software widely used in the aerospace industry to simulate and evaluate the thermal behavior of spacecraft and their components in orbit. By modeling the thermal environment and the heat exchanges between various parts of the spacecraft, Thermica provides accurate predictions of temperature distributions and thermal loads. These simulations are crucial for ensuring that the spacecraft's thermal control systems can maintain operational temperatures within safe limits, thereby protecting sensitive equipment and ensuring mission success.
   \end{itemize}
\begin{figure}[H]
        \centering
        \includegraphics[scale=0.3]{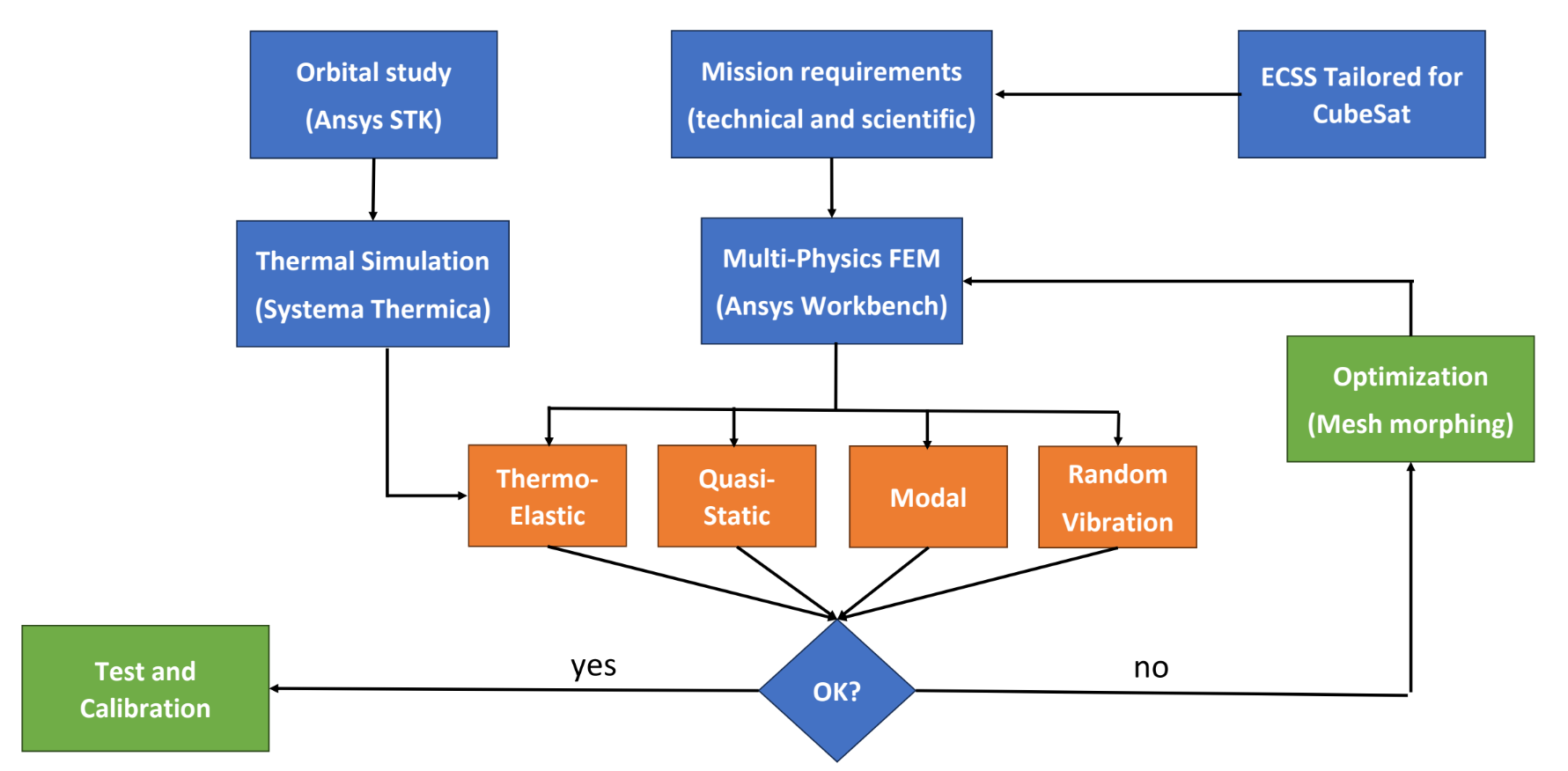}
        \caption{Workflow for method analysis}
        \label{workflow}
\end{figure}
This paper will describe the working method currently being developed, which is intended to be used for making modifications and improvements starting from phase B of the CUSP mission.\\
The evolution of the mission design is based on the implementation of a simplified model, without rounds and attachment points in 2D shell and 3D solid parts, which validates the design choices through a set of numerical multi-physical simulations based on mission requirements and the tailored European ECSS standard \cite{ECSS} for CubeSat missions.\\
Therefore the study was carried out in the following steps:
\begin{itemize}
    \item {detailed analysis of the phase A design, encompassing each component of the assembly and its interface with the platform.}
    \item {obtaining the simplified and defeatured models but compliant in terms of mass, volume and critical part of the structural design on ANSYS SpaceClaim, meshing and import the simplified FEM model in the numerical simulation software.}
   \item {preliminary definition of boundary conditions and constraints for determining multi-physical loads, particularly mechanical and thermal.}
   \item {numerical simulation, validation and control of technical and scientific mission requirement.}
\end{itemize}

\section{PAYLOAD DESIGN}
\label{sec:sections}
\begin{figure}[H]
        \centering
        \includegraphics[scale=0.1]{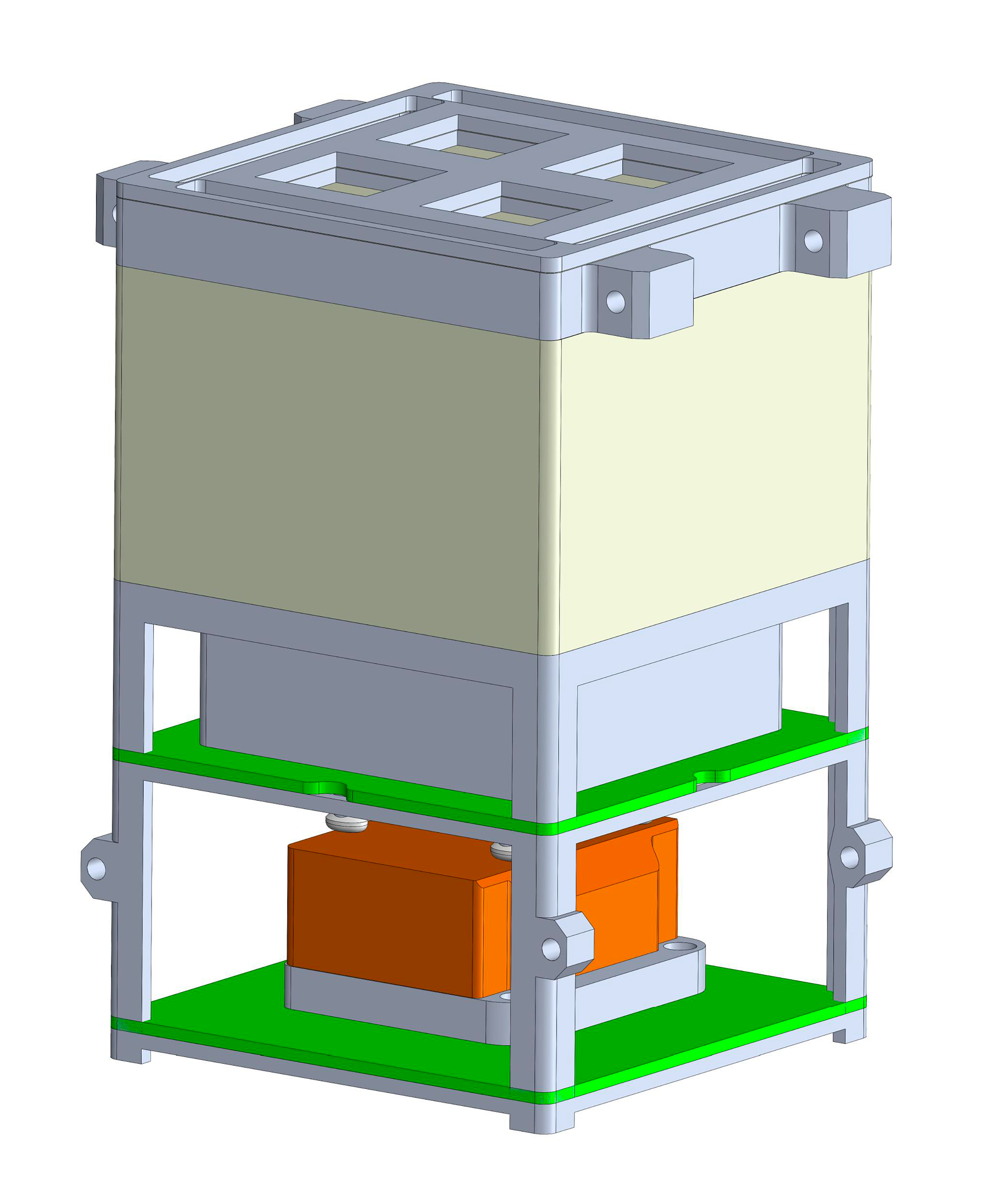}
        \caption{Phase A design.}
        \label{phaseA}
\end{figure}
\begin{figure}[H]
        \centering
        \includegraphics[scale=0.2]{Pictures/cusp1.png}\quad
        \includegraphics[scale=0.2]{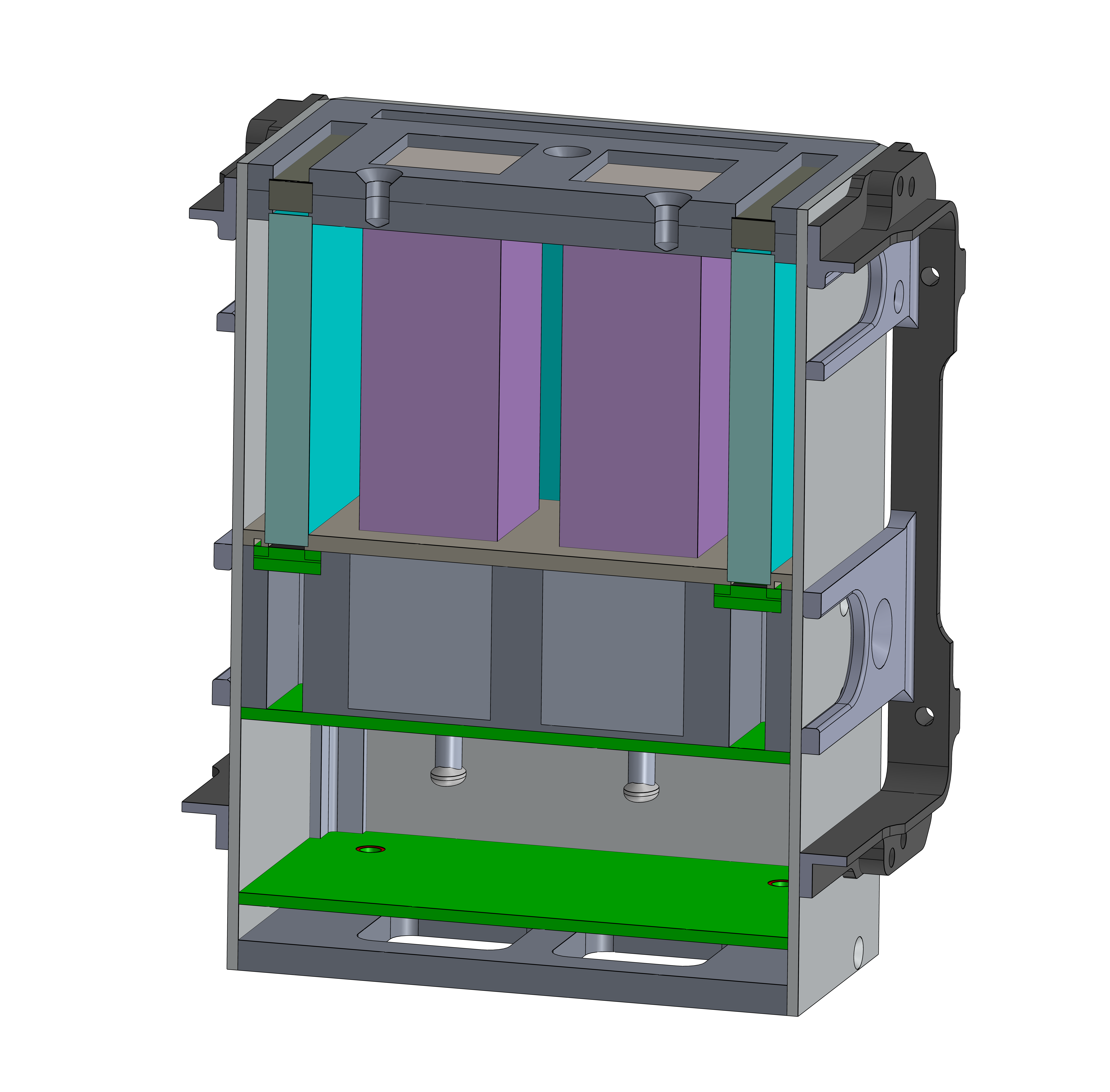}
        \caption{Preliminary payload towards and improved study design.}
        \label{cuspB}
\end{figure}

The mechanical design, from the inception of Phase A of the mission in figure \ref{phaseA}, was already represented in detail since the sensors and sensitive elements of the payload had already been identified.\\
After completion of Phase A some design improvements are under analysis while waiting for the start of Phase B, a review is reported in figure \ref{cuspB}. Redout sensors (photomultiplier tubes and Avalanche photodiode) were already selected, thus this analysis coincided with an in-depth examination of the individual mechanical components, the interfaces among these components, and their integration with the overall platform.
The payload consist of:
\begin{itemize}
    \item {Two identical aluminium interfaces with the platform;}
    \item {An aluminium alloy top, bottom lid and 4x panels with a passive tungsten film clamped;}
   \item {The collimator tray that contain 4 collimators and filters for scatterer and 4 collimators and filters for absorber;}   
   \item {32x absorber made from GAGG scintillator hosted on APD board;}   
   \item {In the same mechanical and electrical framework of the scintillator tray are located 64x scatterer made from plastic scintillator and leaned on the Detector frame;}
    \item {A set of 4x ribs contributing to the mechanical connection of the parts.}
\end{itemize}
\section{MULTI-PHYSICS FEM}
\begin{figure}[H]
        \centering
        \includegraphics[scale=0.25]{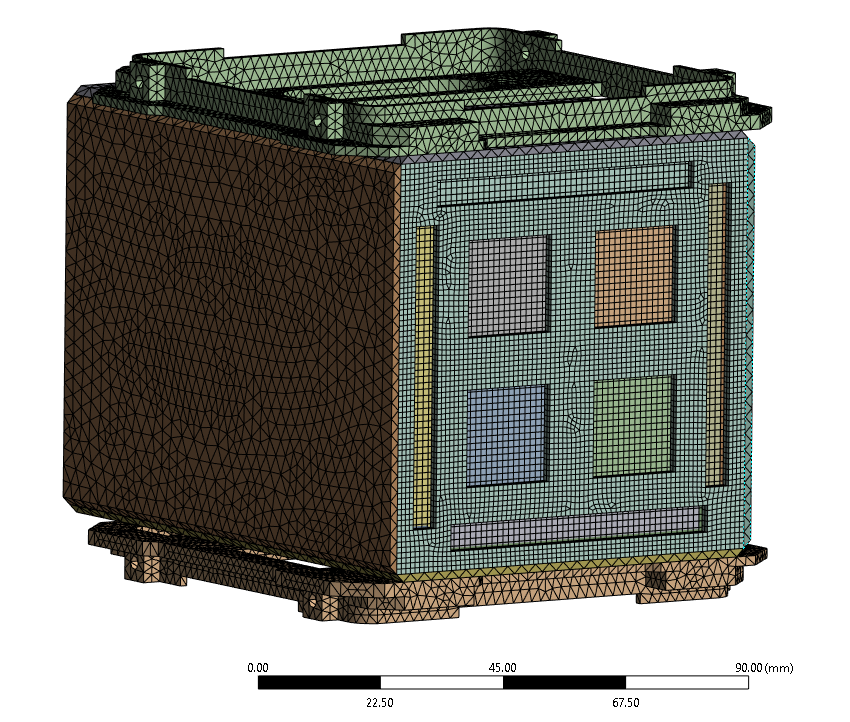}
        \caption{Detailed mesh on payload.}
        \label{mesh}
\end{figure}

The preliminary multiphysics simulation work was conducted using numerical analysis with ANSYS software, specifically ANSYS Workbench. This involved creating a simplified model that maintains the overall envelope and mass of the system while reducing geometric complexity and non$-$essential details.
A simplified model is crucial at this stage to ensure reasonable computation times without sacrificing simulation accuracy. The model includes all major structural and thermal characteristics, ensuring reliable analysis of multiphysical loads.\\
A key aspect of this work was setting up a fine mesh, figure \ref{mesh}, a detailed mesh is essential for accurately capturing physical phenomena such as mechanical deformations and temperature distributions. The fine mesh provides high$-$resolution numerical solutions, ensuring precise and reliable results.
The mesh setup involved selecting the element type, distributing mesh density in critical regions, and optimizing mesh quality to avoid distortions. The combination of a simplified yet accurate model and a fine mesh enabled effective preliminary analysis, providing a solid foundation for further detailed studies and design improvements.\\
A specific material is attributed to each component of the model setting its mechanical properties; the material is taken from the ANSYS library or created by inserting the values of its mechanical properties.\\
Each element that makes up the module is subject to different physical and thermal loads and at the same time will have its own function within the experiment. In table \ref{tab:material} the database of materials created for this study.
\begin{table}[H]
\centering
\begin{tabular}{|>{\centering\arraybackslash}m{2cm}|>{\centering\arraybackslash}m{1,5cm}|>{\centering\arraybackslash}m{2cm}|>{\centering\arraybackslash}m{1,5cm}|>{\centering\arraybackslash}m{2cm}|>{\centering\arraybackslash}m{1,5cm}|>{\centering\arraybackslash}m{1,5cm}|}
\hline
\textbf{Material} & \textbf{Density (kg/m$^3$)} & \textbf{Young's Modulus (GPa)} & \textbf{Poisson Ratio} & \textbf{Thermal Expansion Coefficient (10$^{-6}$/K)} & \textbf{Yield Strength (MPa)} & \textbf{Ultimate Strength (MPa)} \\
\hline
Aluminum 6082 & 2700 & 70 & 0.33 & 23.5 & 260 & 310 \\
\hline
Aluminum 7075 & 2810 & 71.7 & 0.33 & 23.6 & 503 & 572 \\
\hline
Ti6Al4V & 4430 & 113.8 & 0.34 & 8.6 & 880 & 950 \\
\hline
3D Printed PEEK & 1320 & 3.6 & 0.36 & 47 & 90 & 100 \\
\hline
Tungsten & 19300 & 400 & 0.28 & 4.5 & 550 & 750 \\
\hline
Scatterer & 1023 & 1.86 & 0.41 & 80 & 28 & 29.2 \\
\hline
Scintillator GAGG & 6630 & 1.86 & 0.4 & 75 & 28 & 29.2 \\
\hline
\end{tabular}
\caption{Material properties}
\label{tab:material}
\end{table}
Several numerical analyses were conducted to ensure the structural integrity and performance of the payload under various conditions. These analyses included:
\begin{itemize}
    \item Quasi$-$static analysis with loads of 15g per axis to simulate extreme mechanical stress scenarios.
    \item Random vibration analysis using an interpolated curve derived from both European (ECSS) and American (GEVS) standards, ensuring compliance with relevant guidelines and launch stresses.
    \item Thermo$-$elastic analysis incorporating thermal loads obtained from orbital simulations performed using Thermica, to evaluate the payload's response to thermal stresses in orbit.
    \item Modal analysis to verify that the first resonance mode did not exceed 120 Hz, a critical mission requirement (100 Hz plus margin).
\end{itemize}
These comprehensive analyses provided a robust understanding of the payload's behavior under the expected operational conditions.

\section{RESULTS AND PROPOSAL}
\begin{figure}[H]
        \centering
        \includegraphics[scale=0.5]{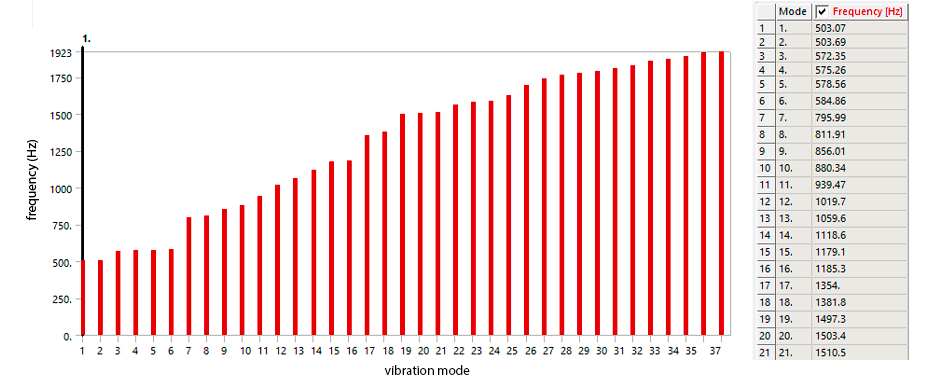}
        \caption{Results for modal analysis.The first six modes are rigid body forms, meaning the rotational and translational oscillations of the entire system. All subsequent modes represent the elastic forms.}
        \label{modal}
\end{figure}

\begin{figure}[H]
        \centering
        \includegraphics[scale=0.28]{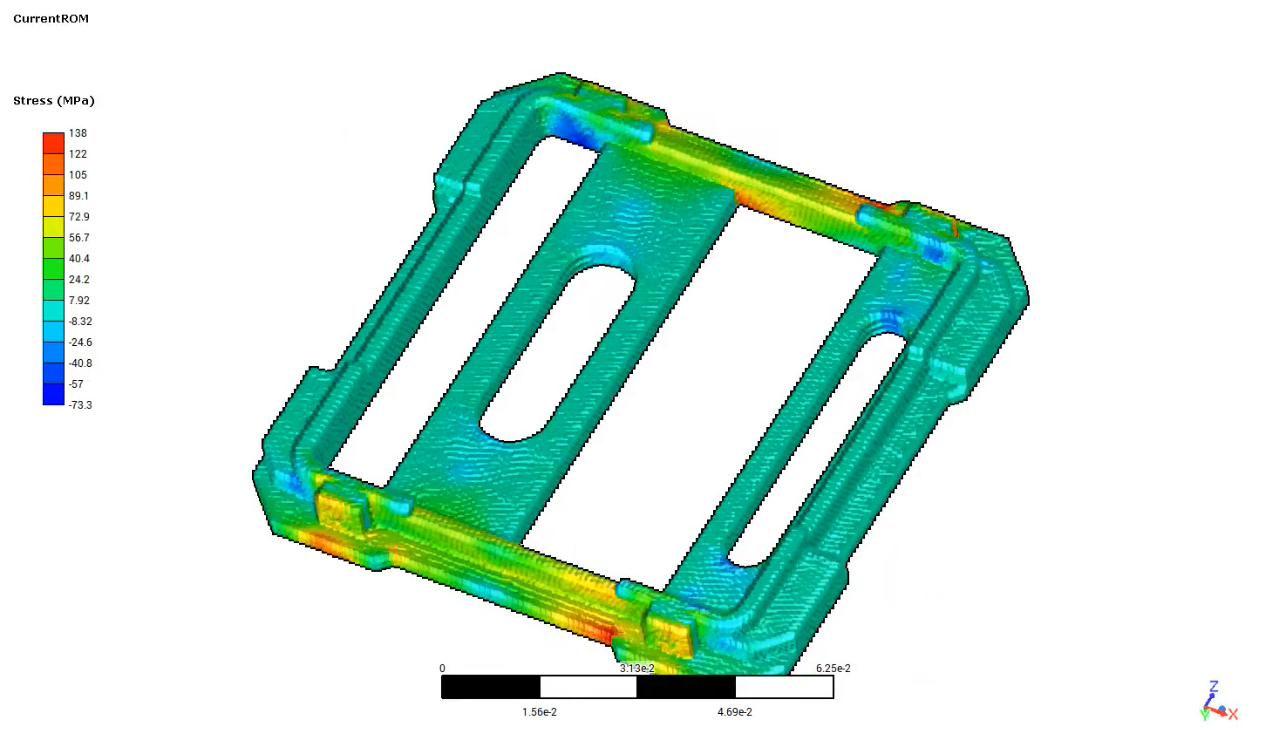}
        \caption{Rom builder for the I/f payload-platform.}
        \label{rom1}
\end{figure}

\begin{figure}[H]
        \centering
        \includegraphics[scale=0.23]{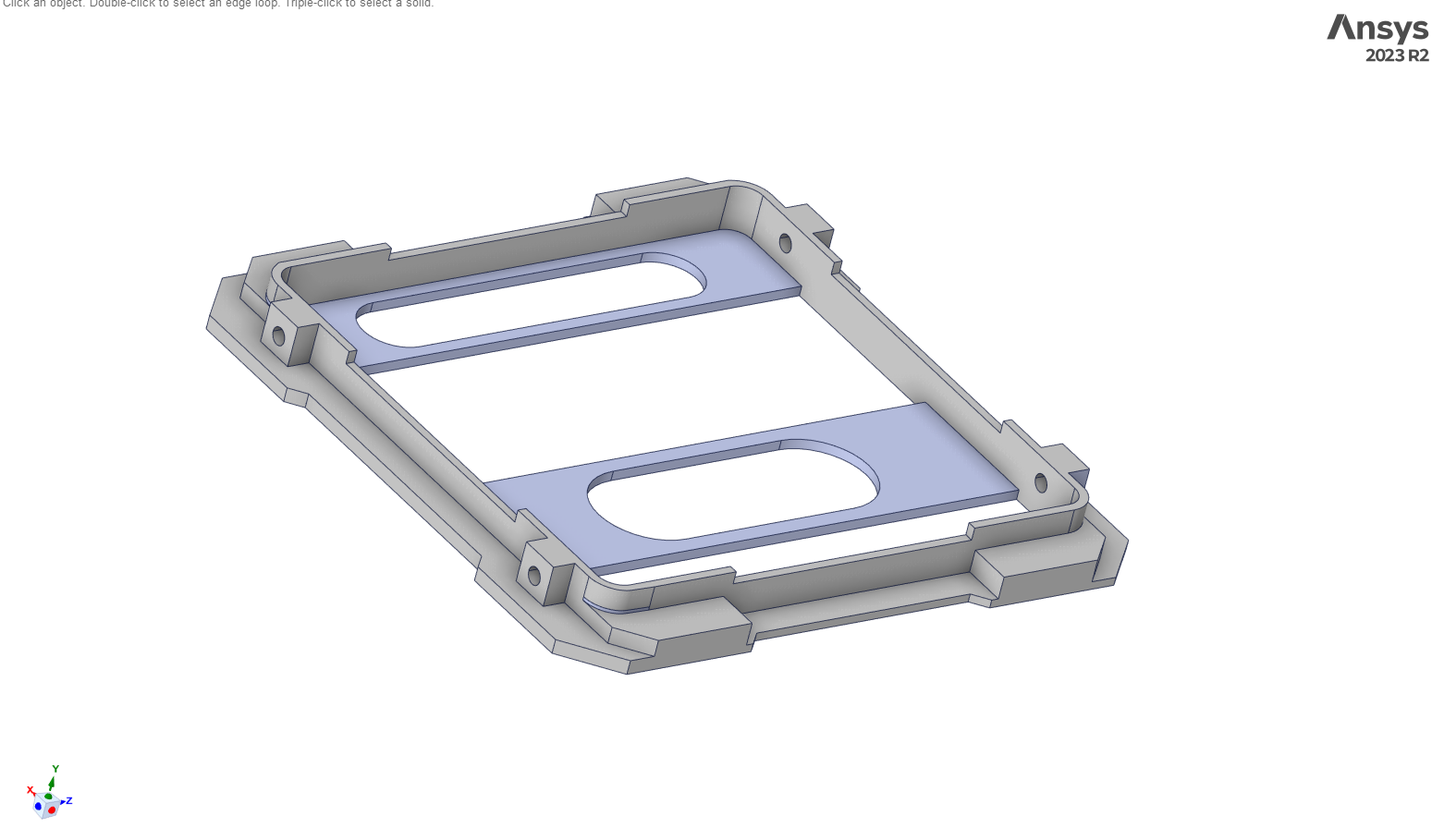}
        \caption{Preliminary optimization for the interfaces.}
        \label{geom1}
\end{figure}

\begin{figure}[H]
        \centering
        \includegraphics[scale=0.3]{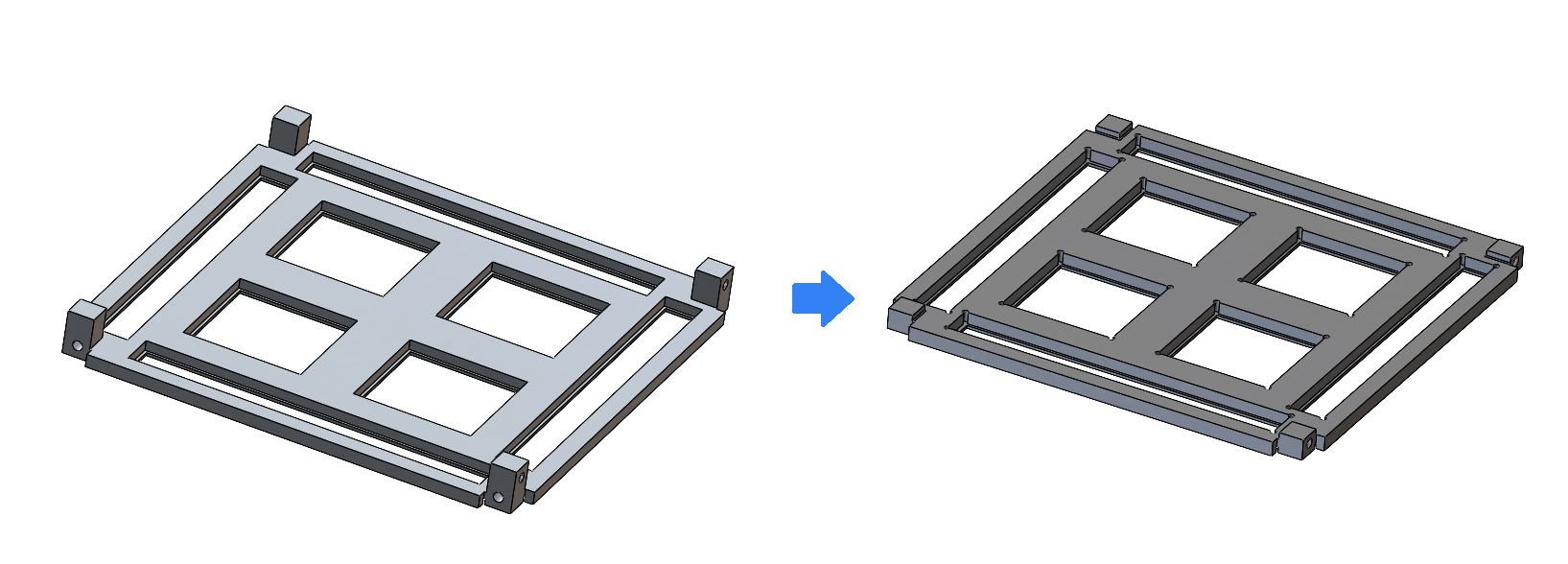}
        \caption{Preliminary optimization for the scintillator frame.}
        \label{opti1}
\end{figure}

\begin{figure}[H]
        \centering
        \includegraphics[scale=0.13]{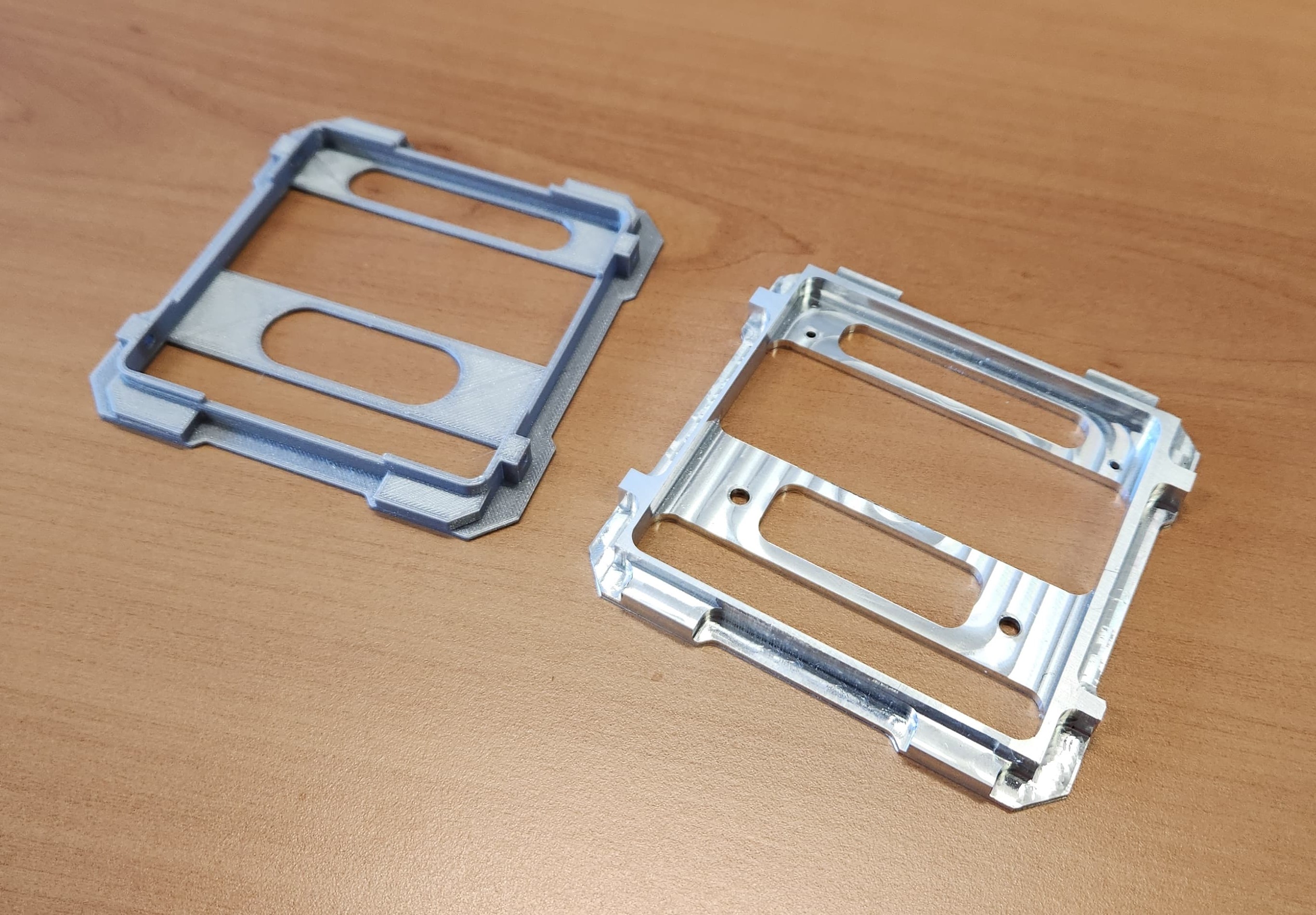}\
        \caption{Technological demonstrator.}
        \label{manu2}
\end{figure}

The numerical simulations conducted on the CUSP payload yielded highly positive results for both the modal and quasi-static analyses on each axis. The quasi-static analysis, which applied loads of 15g per axis to simulate extreme mechanical stress scenarios, confirmed the structural integrity of the payload under these conditions. Similarly, the modal analysis, aimed at verifying that the first resonance mode exceed the mission requirement of 120 Hz, indicated that the payload's design was well within acceptable limits, in figure \ref{modal}.
The random vibration and thermo-elastic analyses highlited the need of a revision of the design of the scintillator frame (due to random vibration) and the interface with the platform (due to thermo-mechanical stress).
Therefore, a preliminary optimization, see figure \ref{rom1}, was developed. We created a reduced mode for the scintillator frame and the interface with the platform. By employing advanced morphing techniques\cite{bianc}, several iterations were performed, resulting in mechanical prototypes that exhibited improved resilience to the identified loads.\\
These optimized components, shown in figure \ref{geom1} and figure \ref{opti1} were subjected to multiple rounds of testing and refinement, ensuring that they met the required performance standards.
Following the completion of these analyses, a technological demonstrator was developed and fabricated in the workshop.
\begin{figure}[H]
        \centering
        \includegraphics[scale=0.2]{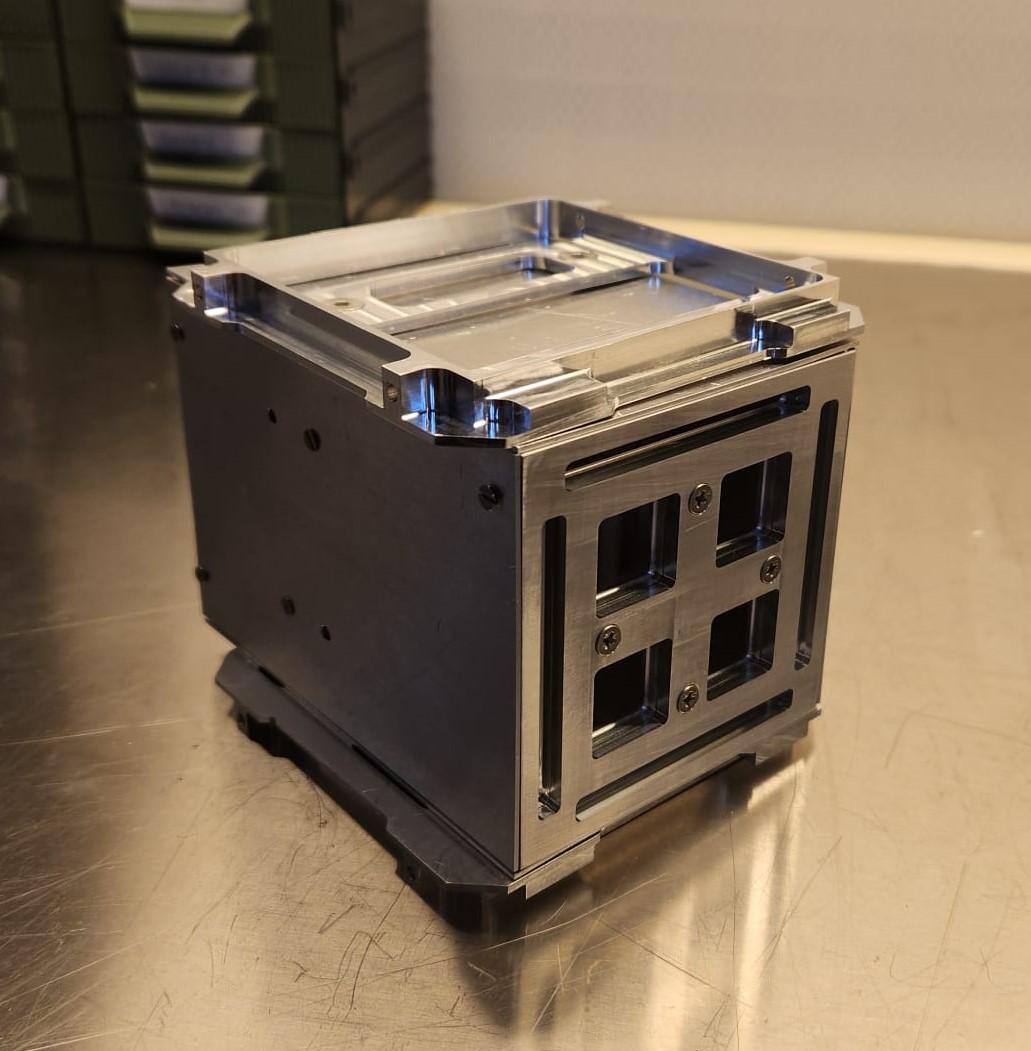}
        \caption{Technological demonstrator.}
        \label{manu1}
\end{figure}
A demonstrator like this, in figure \ref{manu1} incorporating the optimized components, will be tested in the subsequent phases of the mission to validate its performance under real-world conditions. The parts that underwent optimization and were produced using the outlined method were 3D printed, see figure \ref{manu2} leveraging the precision and flexibility of additive manufacturing to achieve the desired specifications.

\section{CONCLUSION}
The comprehensive approach discussed in this paper allowed to improve the design of the payload by increasing he robustness of its components. This iterative process of analysis, optimization, and fabrication enhanced the overall reliability and effectiveness of the CUSP payload, ensuring its capability to withstand the rigors of space operations.\\
The implemented method has proven to be highly effective in identifying and mitigating potential weaknesses in the payload design. Moving forward, we plan to incorporate this method into the comprehensive multi physics design of the mission payload. By doing so, we aim to achieve a higher level of accuracy and reliability, ultimately ensuring the payload's successful operation in the challenging conditions of space.\\
In conclusion, future steps for the project involve more detailed numerical analyses followed by additional enhancement, taking into account all manufacturing processes and their inherent
limitations. Finally, to validate the generated model, environmental tests will be performed on a technological demonstrator.

\bibliography{main} 
\bibliographystyle{spiebib} 

\acknowledgments 
Activity funded by ASI phase A contract 2022-4-R.0.

\end{document}